\documentclass[aps,graphicx,amsmath,amssymb,onecolumn]{revtex4}

\usepackage{graphicx}
\usepackage{dcolumn}
\usepackage{bm}
\usepackage{booktabs}
\usepackage{CJK}
\usepackage{subfigure}
\usepackage{epsfig}

\begin{document}

\title{Violation of generalized Bell inequality and its optimal measurement settings}

\author{Dong Ding$^{1}$, Yingqiu He$^{2}$}
\email {yingqiuhe@126.com}
\author{Fengli Yan$^3$}
 \email{flyan@hebtu.edu.cn}
\author{Ting Gao$^4$ }
 \email{gaoting@hebtu.edu.cn}
\affiliation {$^1$ College of Science, North China Institute of Science and Technology, Beijing 101601, China \\
$^2$ Department of Biomedical Engineering, Chengde Medical University, Chengde 067000, China \\
$^3$ College of Physics and Information Engineering, Hebei Advanced Thin Film Laboratory, Hebei Normal University, Shijiazhuang 050024, China\\
$^4$ College of Mathematics and Information Science, Hebei Normal University, Shijiazhuang 050024, China}
\date{\today}

\begin{abstract}

We provide  a method to describe  quantum nonlocality for $n$-qubit systems.
By treating the correlation function as an $n$-index tensor, we derive a generalized Bell inequality.
Taking generalized Greenberger-Horne-Zeilinger (GHZ) state for example, we calculate quantum prediction under a series of measurement settings involving various angle parameters.
We reveal the exact relationship between quantum prediction and the angle parameters. We show that there exists a set of optimal measurement settings and find the corresponding maximal quantum prediction  for $n$-qubit generalized  GHZ states. As an example, we consider an interesting situation involving only two angle parameters. Finally, we obtain a criterion for the violation of the generalized Bell inequality.

\end{abstract}

\maketitle

\section{Introduction}

To demonstrate the nonlocal quantum correlation of quantum system, in 1964, Bell \cite{Bell1964} proved that quantum predictions are incompatible with the local
hidden variable (LHV) model by a simple logical contradiction.
Inspired by this seminal paper, Clauser \emph{et al.} \cite{CHSH1969} derived a correlation inequality, namely Clauser-Horne-Shimony-Holt (CHSH) inequality,
which provides a way of experimentally testing the LHV theory.
Then, a series of multipartite Bell-type inequalities have been proposed
\cite{Mermin1990,Ardehali1992,BK1993, WW2001,ZB2002, WYKO2007PRA75-032332, GSDRS2009, LF2012, WZCG2013, DHYG2015CPB, HDYG2015EPL}, where Werner-Wolf-\.{Z}ukowski-Brukner (WWZB) inequalities \cite{WW2001,ZB2002} are the most important because of the properties of investigating the possible connections between quantum nonlocality and entanglement for $n$-qubit systems.
For two-qubit systems, the theorem of Gisin \cite{G91, GP92} states that all pure entangled states violate the CHSH inequality.
Then, Chen \emph{et al.} \cite{CWKO2004} generalized Gisin's theorem to three-qubit system and showed that all three-qubit generalized Greenberger-Horne-Zeilinger (GHZ) states violate a Bell inequality for probabilities.
On the other hand, given a set of standard Bell experiment settings involving two dichotomic observables for each position, there exist pure $n$-qubit entangled states that do not violate any Bell inequality \cite{ZBLW2002}.
By now it is an open question whether Gisin's theorem can be generalized to an arbitrary $n$-qubit system \cite{BCPSW2014}.
In any event, the key to investigating Bell inequalities with multipartite correlation functions is to find a set of optimal measurement settings.

Optical quantum systems \cite{DDI2006, Kok2007, Pan2012, LDJW2007, AGACVMC2012,CM2014,LDQ2015,KNM2017} are prominent candidates for testing nonlocal quantum correlations, since multiphoton entanglement, interferometry and measurement are relatively easy to perform in experiments, as long as the number of photons is not very large.
In 2001, Weinfurter and \.{Z}ukowski  \cite{WZ2001} proposed a scheme to produce a  superposition of four-photon GHZ state with a product state of two Bell states and investigated the features of this state  by constructing a Bell-type inequality.
By introducing a set of polarization correlation measurements, it has been shown that with a given set of measurement settings the maximal violation of the Bell-type inequality is $4\sqrt{2}/3$.
Later, with this method, Li and Kobayashi \cite{LK2004} investigated another four-photon superposition state,
and more recently, Ding \emph{et al.} \cite{DHYG2017-4-photon} discussed a class of generic superposition of four-photon entangled states with a tunable angle parameter.
Of course, a natural question is whether or not the given settings \cite{WZ2001, LK2004, DHYG2017-4-photon} are optimal or unique.

In this paper, we investigate  quantum nonlocality for $n$-qubit entangled states.
We first describe a method of treating correlation function as an $n$-index tensor and then derive a generalized Bell inequality.
Under a set of measurement settings involving various angle parameters, we calculate quantum prediction of generalized GHZ states and show the exact relationship between quantum prediction and the angle parameters.
We demonstrate  that there is a set of optimal measurement settings and  obtain the corresponding maximal quantum prediction  for $n$-qubit generalized  GHZ states. We analyze the  interesting situation involving only two angle parameters in details. Finally, as an important application,  a  criterion for the violation of the generalized Bell inequality is provided.

\section{A generalized Bell inequality for $n$-qubit system}

In LHV theory \cite{BCPSW2014}, a correlation function represents an average over many runs of experiment.
An $n$-partite correlation function for two alternative dichotomic measurements is generally given by
\begin{eqnarray}\label{E-LHV}
E_{\textrm{LHV}}(\phi_1^{k_1},\phi_2^{k_2},\cdots,\phi_n^{k_n})
= \int d\lambda \rho(\lambda) I_{1}(\phi_1^{k_1}, \lambda) I_{2}(\phi_2^{k_2}, \lambda) \cdots I_{n}(\phi_n^{k_n}, \lambda),
\end{eqnarray}
where  $\lambda$ is a hidden variable and $\rho(\lambda)$ is the probability distribution function, $\phi_i^{k_i}$, $(k_i=0,1)$ indicates the local phase angle at site $i$, and $I_{i}(\phi_i^{k_i}, \lambda)=\pm 1$, $i=1,2,\cdots,n$, represents the predetermined binary outcomes of the measurements.
As described by Weinfurter and \.{Z}ukowski  \cite{WZ2001},
one can treat the four-photon correlation function as a four-index tensor.
Here we consider an $n$-index tensor which is obtained by taking the form of a tensor product of $n$ two-dimensional real vectors $\textbf{v}_{i}^{\lambda}=(I_{i}(\phi_{i}^{0}, \lambda),I_{i}(\phi_{i}^{1}, \lambda))$, $i=1,2,\cdots,n$.
That is, we define a tensor for $n$-partite system as
\begin{eqnarray}\label{E-LHV-lambda}
\hat{E}_{\textrm{LHV}}
= \int d\lambda \rho(\lambda) \textbf{v}_{1}^{\lambda} \otimes \textbf{v}_{2}^{\lambda} \otimes \cdots \otimes \textbf{v}_{n}^{\lambda}.
\end{eqnarray}
Choose two orthogonal unit vectors
\begin{eqnarray}\label{}
\textbf{v}_{i}^{0}=(1,0),~~~~ \textbf{v}_{i}^{1}=(0,1).
\end{eqnarray}
Since each of the vectors $\textbf{v}_{i}^{\lambda}$ can be written as $\sum_{k_{i}=0,1}I_{i}(\phi_{i}^{k_{i}}, \lambda)\textbf{v}_{i}^{k_{i}}$, one can  simplify the correlation function as
\begin{eqnarray}\label{}
\hat{E}_{\textrm{LHV}} &=&
\sum_{k_1,k_2,\cdots,k_n = 0,1} E_{\textrm{LHV}}(\phi_1^{k_1},\phi_2^{k_2},\cdots,\phi_n^{k_n}) \textbf{v}_{1}^{k_1} \otimes \textbf{v}_{2}^{k_2} \otimes \cdots \otimes \textbf{v}_{n}^{k_n}.
\end{eqnarray}

Let $\textbf{A}_{i}^{0}=(1,1)$ and $\textbf{A}_{i}^{1}=(1,-1)$.
Obviously,
\begin{eqnarray}\label{}
\textbf{v}_{i}^{k_i} = \frac{1}{2}[\textbf{A}_{i}^{0}+(-1)^{k_i}\textbf{A}_{i}^{1}], ~~~ k_{i}=0,1, ~~~ i=1,2,\cdots,n.
\end{eqnarray}
The correlation function can be expressed as
\begin{eqnarray}\label{}
\hat{E}_{\textrm{LHV}} &=&
\frac{1}{2^{n}} \sum_{j_1,j_2,\cdots,j_n = 0,1} \sum_{k_1,k_2,\cdots,k_n = 0,1} (-1)^{\textbf{k} \centerdot \textbf{j}}  \nonumber \\
& & \times E_{\textrm{LHV}}(\phi_1^{k_1},\phi_2^{k_2},\cdots,\phi_n^{k_n})
\textbf{A}_{1}^{j_{1}}  \otimes \textbf{A}_{2}^{j_{2}} \otimes \cdots \otimes \textbf{A}_{n}^{j_{n}},
\end{eqnarray}
where $\textbf{k}=(k_1,k_2,\cdots,k_n)$ and $\textbf{j}=(j_1,j_2,\cdots,j_n)$.
Let
\begin{eqnarray}\label{}
c_{j_1,j_2,\cdots,j_n} = \frac{1}{2^{n}}  \sum_{k_1,k_2,\cdots,k_n = 0,1} (-1)^{\textbf{k} \centerdot \textbf{j}} E_{\textrm{LHV}}(\phi_1^{k_1},\phi_2^{k_2},\cdots,\phi_n^{k_n})
\end{eqnarray}
be the LHV correlation coefficients.
Then we have
\begin{eqnarray}\label{}
\hat{E}_{\textrm{LHV}} &=&
\sum_{j_1,j_2,\cdots,j_n = 0,1} c_{j_1,j_2,\cdots,j_n}
\textbf{A}_{1}^{j_{1}}  \otimes \textbf{A}_{2}^{j_{2}} \otimes \cdots \otimes \textbf{A}_{n}^{j_{n}}.
\end{eqnarray}

Let $\textbf{A}_{i}^{2}=(-1,-1)$ and $\textbf{A}_{i}^{3}=(-1,1)$, then we can use the probability to describe  $n$-partite correlation function (2) as
\begin{equation}\label{}
\hat{E}_{\textrm{LHV}} =
\sum_{j_1,j_2,\cdots,j_n = 0,1,2,3} p_{j_1,j_2,\cdots,j_n}
\textbf{A}_{1}^{j_{1}}  \otimes \textbf{A}_{2}^{j_{2}} \otimes \cdots \otimes \textbf{A}_{n}^{j_{n}},
\end{equation}
where  $p_{j_1,j_2,\cdots,j_n}$ is the probability of  $\textbf{v}_{i}^{\lambda}$ being $\textbf {A}_i^{j_i}, i=1,2, \cdots, n, j_i=0,1,2,3$, i.e.,  $p_{j_1,j_2,\cdots,j_n}$ is the probability of obtaining measurement outcomes $\textbf {A}_1^{j_1}, \textbf {A}_2^{j_2}, \cdots, \textbf {A}_n^{j_n}$.
Obviously,
 $$\sum_{j_1,j_2,\cdots,j_n = 0,1,2,3}p_{j_1,j_2,\cdots,j_n}=1.$$
Note that $\textbf A_i^{j_i+2}=-\textbf A_i^{j_i}$ with $j_i=0,1$, we obtain
\begin{eqnarray}\label{}
\hat{E}_{\textrm{LHV}} &=&
\sum_{j_1,j_2,\cdots,j_n = 0,1} (p_{j_1,j_2,\cdots,j_n}-p_{j_1+2,j_2,\cdots,j_n} - \cdots - p_{j_1,j_2,\cdots,j_{n-1},j_n+2}\nonumber\\
                    & &+p_{j_1+2,j_2+2,j_3,\cdots,j_n} + \cdots - p_{j_1+2,j_2+2,j_3+2,j_4,\cdots,j_n} - \cdots)  \textbf{A}_{1}^{j_{1}}\otimes \textbf{A}_{2}^{j_{2}} \otimes \cdots \otimes \textbf{A}_{n}^{j_{n}}.
\end{eqnarray}
Compared with Eq.(8), one gets
\begin{eqnarray}\label{}
&c_{j_1,j_2,\cdots,j_n}=&p_{j_1,j_2,\cdots,j_n}-p_{j_1+2,j_2,\cdots,j_n} - \cdots - p_{j_1,j_2,\cdots,j_{n-1},j_n+2}
                    +p_{j_1+2,j_2+2,j_3,\cdots,j_n} + \cdots \nonumber\\
                    &&- p_{j_1+2,j_2+2,j_3+2,j_4,\cdots,j_n} - \cdots.
\end{eqnarray}
Then
\begin{eqnarray}\label{BI-LHV}
\sum_{j_1,j_2,\cdots,j_n = 0,1} |c_{j_1,j_2,\cdots,j_n}| \leq 1.
\end{eqnarray}
This inequality is derived from the  natural generalization of the four-qubit correlation inequality \cite{WZ2001}  to $n$-qubit systems. It is conventionally referred to as generalized Bell inequality, and can be used to test the LHV theory. It may be  equivalent to WWZB inequality \cite {WW2001, ZB2002} and  limits the total amount of correlation allowed for the LHV theory.

On the other hand, quantum mechanically, suppose a measurement, described by measurement operator
\begin{equation}\label{}
\{M_{x} = |m_{x},\phi_{x}\rangle \langle m_{x},\phi_{x}|, m_x=\pm 1\}
\end{equation}
is performed upon $x$-port with a detector placed at the corresponding output station,
where
\begin{equation}\label{}
|m_{x},\phi_{x}\rangle = \frac{1}{\sqrt{2}}(|0\rangle_{x} + m_{x} \text{e}^{-\text{i}\phi_{x}} |1\rangle_{x}), ~~~ x=1,2,\cdots, n,
\end{equation}
$\phi_x$ is a local phase setting chosen by each of the observers and $m_x$ represents the possible measurement result.

Consider a standard quantum correlation test, in which each observer chooses between two dichotomic measurements;
that is, for each site one can label phase angle $\phi_x^{k_x}$, $k_x=0,1$ and take $m_x=\pm 1$.
For an $n$-qubit entangled state $|\psi_n \rangle$, the probability of outcomes $m_{1},m_{2}, \cdots,  m_{n}$ with the phase settings $\phi_1^{k_1},\phi_2^{k_2},\cdots,\phi_n^{k_n}$ labels $p(m_1,m_2,\cdots,m_n|\phi_1^{k_1},\phi_2^{k_2},\cdots,\phi_n^{k_n})$
and then the  correlation function can be represented by
\begin{eqnarray}\label{n-QM-correlation}
E_{\textrm{QM}}(\phi_1^{k_1},\phi_2^{k_2},\cdots,\phi_n^{k_n}) &=& \sum_{m_1,m_2,\cdots,m_n = \pm 1} p(m_1,m_2,\cdots,m_n|\phi_1^{k_1},\phi_2^{k_2},\cdots,\phi_n^{k_n}) m_1m_2\cdots m_n,
\end{eqnarray}
where the sum is over all possible runs of experiment.
This set of operators is sufficient to describe the quantum correlation in contrast to the LHV case.

Similarly, the quantum correlation function can also be described by the $n$-fold tensor
\begin{eqnarray}\label{}
\hat{E}_{\textrm{QM}} &=&
\sum_{j_1,j_2,\cdots,j_n = 0,1} q_{j_1,j_2,\cdots,j_n}
\textbf{A}_{1}^{j_{1}}  \otimes \textbf{A}_{2}^{j_{2}} \otimes \cdots \otimes \textbf{A}_{n}^{j_{n}},
\end{eqnarray}
where
\begin{eqnarray}\label{}
q_{j_1,j_2,\cdots,j_n} = \frac{1}{2^{n}}  \sum_{k_1,k_2,\cdots,k_n = 0,1} (-1)^{\textbf{k} \centerdot \textbf{j}} E_{\textrm{QM}}(\phi_1^{k_1},\phi_2^{k_2},\cdots,\phi_n^{k_n})
\end{eqnarray}
are quantum correlation coefficients.
Compared with the inequality (\ref{BI-LHV}) derived from the LHV correlation,  once quantum prediction
\begin{eqnarray}\label{quantum-prediction}
\sum_{j_1,j_2,\cdots,j_n = 0,1} |q_{j_1,j_2,\cdots,j_n}|
\end{eqnarray}
is greater than the classical limit value 1, which means that Bell inequality is violated and quantum nonlocal correlations of quantum system occurs.

\section{Quantum prediction for $n$-qubit generalized GHZ states}

In a finite-dimensional Hilbert space, consider an  $n$-qubit generalized GHZ state
\begin{eqnarray}\label{GGHZ}
 |\psi_{n}\rangle &=& \alpha |00 \cdots 0\rangle_{12 \cdots n} + \beta |11 \cdots 1\rangle_{12 \cdots n},
\end{eqnarray}
where $\alpha$ and $\beta$ are respectively the complex parameters satisfying the normalization condition $|\alpha|^{2}+|\beta|^{2}=1$.
A computation reveals  that the quantum correlation function determined by the dichotomic measurement parameter settings $\{\phi_1^{k_1},\phi_2^{k_2},\cdots,\phi_n^{k_n}\}$ for the generalized GHZ state (19) is
\begin{eqnarray}\label{}
E_{\textrm{QM}}(\phi_1^{k_1},\phi_2^{k_2},\cdots,\phi_n^{k_n}) &=& 2 \text{Re}[\alpha\beta^{*}\text{e}^{-\text{i}(\phi_1^{k_1}+\phi_2^{k_2}+\cdots+\phi_n^{k_n})}].
\end{eqnarray}
Without loss of generality we suppose that $\alpha$ and $\beta$ are real.
Then the quantum prediction is
\begin{eqnarray}\label{sum-QC-1}
\sum_{j_1,j_2,\cdots,j_n = 0,1} |q_{j_1,j_2,\cdots,j_n}| =
\frac{|\alpha\beta^{}|}{2^{n-1}} \sum_{j_1,j_2,\cdots,j_n = 0,1} |\sum_{k_1,k_2,\cdots,k_n = 0,1} (-1)^{\textbf{k} \centerdot \textbf{j}}
\text{Re}[\text{e}^{-\text{i}(\phi_1^{k_1}+\phi_2^{k_2}+\cdots+\phi_n^{k_n})}]|.
\end{eqnarray}

Let
\begin{equation}
\alpha_l=\frac {(\phi_l^1+\phi_l^0)}{2}, ~~~~~ \beta_l=\frac {(\phi_l^1-\phi_l^0)}{2}, ~~~~~ l=1,2,\cdots,n.
\end{equation}
We obtain the the quantum prediction
\begin{eqnarray}\label{sum-QC-1*}
&&\sum_{j_1,j_2,\cdots,j_n = 0,1} |q_{j_1,j_2,\cdots,j_n}|\nonumber\\
&=&|\alpha\beta|\{[|\cos(\sum_{l=1,2,\cdots,n}\alpha_l)|+|\sin(\sum_{l=1,2,\cdots,n}\alpha_l)|]\prod_{l=1,2,\cdots,n}(|\cos\beta_l|+|\sin\beta_l|)\nonumber\\
 & &+[|\cos(\sum_{l=1,2,\cdots,n}\alpha_l)|-|\sin(\sum_{l=1,2,\cdots,n}\alpha_l)|]\prod_{l=1,2,\cdots,n}(|\cos\beta_l|-|\sin\beta_l|)\}.
\end{eqnarray}
The details of derivation can be found in Appendix A.  Eq. (\ref{sum-QC-1*}) is  the exact quantum prediction under a series of measurement settings involving various angle parameters.

Let $\sum_{l=1,2,\cdots,n}\alpha_l=\beta_{0}$. In Appendix B, we prove that
 the maximum value of the quantum prediction for $n$-qubit generalized GHZ state  is
\begin{equation}
\textrm{max}\sum_{j_1,j_2,\cdots,j_n = 0,1} |q_{j_1,j_2,\cdots,j_n}|=|\alpha\beta|
2^{\frac {n+1}{2}},
\end{equation}
which occurs at $\beta_{l}=(2k+1)\pi/4$, $k=0, \pm1,\pm2,\cdots,$ $l=0,1,2,\cdots, n$.

Up to now£¬ we have described the method to investigate the generalized Bell inequality.
One of the attractive aspects of this result is that it provides a convenient way to choose the optimal measurement settings for testing Bell-type inequalities. In fact, according to this result it turns out that the previous settings \cite{WZ2001, LK2004, DHYG2017-4-photon} are optimal, but not unique.

\section{An example }

As an example, we here discuss an interesting situation involving only two angle parameters.
Choose a set of measurement settings satisfying
\begin{eqnarray}\label{}
&\phi_{i}^{0}=0, ~~~ \phi_{i}^{1}=\theta_{1},  ~~~ i=1,2,\cdots,l,\\
&\phi_{i}^{0}=\theta_{2}, ~~~ \phi_{i}^{1}= -\theta_{2}, ~~~ i=l+1,l+2,\cdots,n.
\end{eqnarray}
In this architecture, obviously
\begin{eqnarray}\label{setting-2-angle}
\sum_{i=1,2,\cdots,n}\alpha_i=\frac{l\theta_{1}}{2}, ~~~ \beta_{i=1,2,\cdots,l}=\frac{\theta_{1}}{2},  ~~~ \beta_{i=l+1,l+2,\cdots,n}= -\theta_{2}.
\end{eqnarray}
It is easily seen that for the generalized GHZ state, the quantum prediction
\begin{eqnarray}\label{Q-n-l}
\sum_{j_1,j_2,\cdots,j_n = 0,1} |q_{j_1,j_2,\cdots,j_n}|  &=& |\alpha\beta|
[
 (|\text{cos}{\frac{l\theta_{1}}{2}}|+|\text{sin}{\frac{l\theta_{1}}{2}}|)(|\text{cos}{\frac{\theta_{1}}{2}}|+|\text{sin}{\frac{\theta_{1}}{2}}|)^{l} (|\text{cos}{\theta_{2}}|+|\text{sin}{\theta_{2}}|)^{n-l} \nonumber \\ & &
+(|\text{cos}{\frac{l\theta_{1}}{2}}|-|\text{sin}{\frac{l\theta_{1}}{2}}|)(|\text{cos}{\frac{\theta_{1}}{2}}|-|\text{sin}{\frac{\theta_{1}}{2}}|)^{l}
 (|\text{cos}{\theta_{2}}|-|\text{sin}{\theta_{2}}|)^{n-l}
].
\end{eqnarray}
This implies that the quantum prediction of the given $n$-qubit system varies as two angle parameters $\theta_{1}$, $\theta_{2}$ and the value $l$.

Here we analyze the quantum prediction for the expression (\ref{Q-n-l}).
For the sake of simplicity, we take $\alpha=\beta=1/\sqrt{2}$. As shown in Fig. 1, taking $n=4$ and $l=1$ for example, it is straightforward to show that in the range $0$ to $\pi/2$ the value of quantum prediction varies continuously as angle parameters $\theta_{1}$ and $\theta_{2}$, and the peak $2\sqrt{2}$ occurs at $\theta_1=\pi/2$ and $\theta_2=\pi/4$.
Indeed, similar results can be found for any value of $n$ and $l$.
This allows a further simplification in which one of two angle parameters is fixed, and then one can plot quantum prediction as a function of the remaining phase angle.
In this way, one may, therefore, optimize the measurement settings.
We also take $n=4$ for example. Setting $\theta_{1}=\pi/2$ and taking $\theta=\theta_{2}$, as shown in Fig. 2, this plot shows the quantum prediction varies as the angle parameter $\theta$ over a range from $0$ to $\pi/2$.
By comparing the curves with different values of $l$, one sees immediately that the maximum value $2\sqrt{2}$ occurs at $\theta_{2}=\pi/4$ for $l=1$ and $l=3$, respectively.

To sum up, there are three experimentally significant points in our architecture.
(\expandafter{\romannumeral 1}) The optimal measurement settings are $\theta_1=\pi/2$ and $\theta_2=\pi/4$ with all odd $l$, and the maximum of quantum prediction is $2^{(n-1)/2}$.
(\expandafter{\romannumeral 2}) For $l=0$, it is not the optimal measurement settings and its maximal violation is $2^{(n-2)/2}$ with $\theta_{2}=\pi/4$.
(\expandafter{\romannumeral 3}) For $l=n$, with odd $l$ the maximum of quantum prediction occurs at the optimal measurement settings $\theta_{1}=\pi/2$, while for even $l$ it is not the optimal settings.
In fact, this result is easy to check directly by inserting $\theta_1=\pi/2$ and $\theta_2=\pi/4$ into measurement settings (\ref{setting-2-angle}).
Obviously, with an odd $l$, we have $\sum_{i=1,2,\cdots,n}\alpha_i={(2k+1)\pi}/{4}$, $\beta_{i=1,2,\cdots,l}={\pi}/{4}$ and $\beta_{i=l+1,l+2,\cdots,n}= -{\pi}/{4}$, then the maximum of quantum prediction occurs.

\begin{figure}
  \epsfig{figure=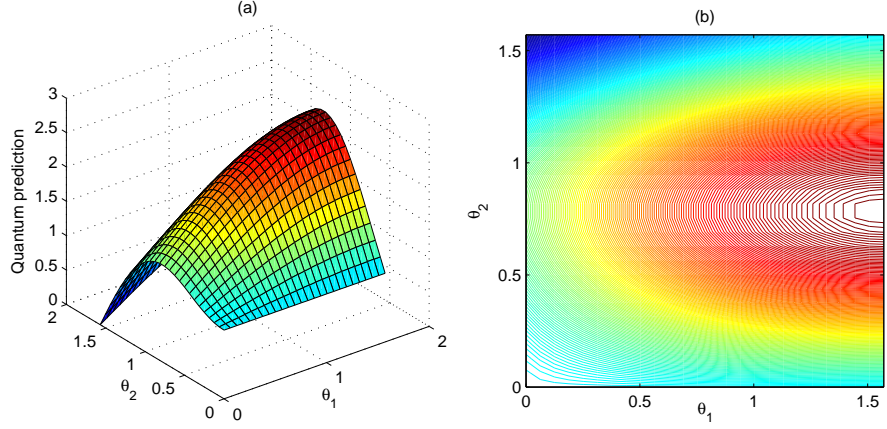,width=0.65\textwidth}
  \caption{(color online). (a) Numerically calculated quantum prediction as a function of $\theta_1$ and $\theta_2$ for four-qubit GHZ state with $l=1$.
(b) The contour lines of (a) show that the maximum occurs at $\theta_1=\pi/2$ and $\theta_2=\pi/4$.}
  \label{}
\end{figure}

\begin{figure}
  \epsfig{figure=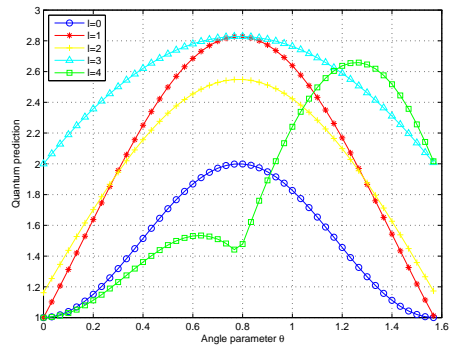,width=0.325\textwidth}
  \caption{(color online). Numerical results of the relationship between quantum prediction and phase angle $\theta$ for the four-qubit GHZ state.}
  \label{}
\end{figure}

\section{Generalization and application}

We now consider the generalized GHZ states with complex coefficients $\alpha$ and $\beta$.
Let $\alpha\beta^{*}=|\alpha\beta^{*}| \text{e}^{-\text{i}\phi}$.
Then, quantum correlation function can be rewritten as
\begin{eqnarray}\label{}
E_{\textrm{QM}}(\phi_1^{k_1},\phi_2^{k_2},\cdots,\phi_n^{k_n}) &=& 2 |\alpha\beta^{*}|\text{cos} (\phi + \phi_1^{k_1}+\phi_2^{k_2}+\cdots+\phi_n^{k_n}).
\end{eqnarray}
A similar calculation yields the quantum prediction
\begin{eqnarray}\label{QP-complex}
\sum_{j_1,j_2,\cdots,j_n = 0,1} |q_{j_1,j_2,\cdots,j_n}|
&=&  |\alpha\beta^{*}| [(|\cos(\phi+\sum_{l=1,2,\cdots,n}\alpha_l)|+|\sin(\phi+\sum_{l=1,2,\cdots,n}\alpha_l)|)\prod_{l=1,2,\cdots,n}(|\cos\beta_l|+|\sin\beta_l|)\nonumber\\
 & & +(|\cos(\phi+\sum_{l=1,2,\cdots,n}\alpha_l)|-|\sin(\phi+\sum_{l=1,2,\cdots,n}\alpha_l)|)\prod_{l=1,2,\cdots,n}(|\cos\beta_l|-|\sin\beta_l|)].
\end{eqnarray}
When $\phi+\sum_{l=1,2,\cdots,n}\alpha_l$ and $\beta_l$ are $(2k+1)\pi/4$ $(k=0,\pm1,\pm2,\cdots)$, the maximum value $|\alpha\beta^{*}| 2^{{(n+1)}/{2}}$ occurs.
According to this result, the generalized Bell inequality will be violated conditioned on
\begin{eqnarray}\label{}
|\alpha\beta^{*}| > 2^{-\frac{n+1}{2}}.
\end{eqnarray}

As an important application, this provides a criterion for the violation of the generalized Bell inequality derived from the LHV theory;
that is, under the present optimal measurement settings,
for $|\alpha\beta^{*}| \leq  2^{-(n+1)/2}$ this inequality can not be violated.
Especially, for real $\alpha$ and $\beta$, if we assume that $\alpha=\text{cos}\xi$ and $\beta=\text{sin}\xi$,
then, consequently, for $\text{sin}2\xi \leq 2^{-(n-1)/2}$ the inequality can not be violated, which is consistent with the result in \cite{ZBLW2002} with an arbitrary odd $n$.

\section{Discussion and summary}

Having investigated the generalized Bell inequality and its maximal violation, we now discuss the possible experimental realization using optical quantum technologies \cite{DDI2006, Kok2007, Pan2012}.
One aspect of  this framework is the preparation of multiphoton entangled states.
With linear optics and multiphoton interferometry, more recently, Pan \emph{et al.} \cite{Pan-8-photon-GHZ-2012, Pan-10-photon-GHZ-2016} successively reported two schemes of observing eight-photon and ten-photon entanglement in experiment.
By combining pairs of photons emitting from the parametric down-conversion processes \cite{SPDC1970, PDC1995}, eight-photon or ten-photon GHZ state can be engineered step by step.
So, with currently available techniques, here one may produce the generalized GHZ state by considering a tunable angle parameter, which is settled by the orientation of a wave plate \cite{WSKPGW2008, LL-NJP2009, DHYG2017-4-photon}.
Another way of preparing multiphoton entangled states is to utilize cross-Kerr nonlinearities \cite{Imoto1985,SI1996,LI2000,NM2004,RV2005,Barrett2005,MNBS2005,Kok2008,HDYG2015OE,HDYG2016SR}.
For example, we may consider an entangler of multiphoton GHZ states proposed by Ding \emph{et al.} \cite{DYG2014}.
By resetting an input state $(\alpha |H\rangle_1+\beta|V\rangle_1)\otimes(|H\rangle_2+|V\rangle_2)\otimes \cdots\otimes(|H\rangle_n+|V\rangle_n)/2^{(n-1)/2}$,
after the homodyne measurement on the probe beam the entangler is capable of preparing an $n$-photon generalized GHZ state.
The other aspect of the present architecture is concerned with performing polarization analysis.
Similar to the four-photon entanglement experiment \cite{EGBKZW2003, GBEKW2003, BEKGWGHBLS2004, XLG2006}, polarization analysis in various bases can be performed in each of the $n$ outputs via quarter- and half-wave plates in front of polarizing beam splitters.
Taking the settings (\ref{setting-2-angle}) (with $l=1$, $\theta_1=\pi/2$ and $\theta_2=\pi/4$) for example,
the observer at site 1 switches analysis angle between 0 and $\pi/2$, and the other observers at sites $2,3,\cdots,n$ switch analysis angles between $\pm \pi/4$. When the $n$ photons are detected by single photon avalanche detectors, an $n$-fold coincidence detection can be registered.
With these registrations one can investigate the violation of the generalized Bell inequality.

In summary, we have shown a method to deal with quantum nonlocality for $n$-qubit systems.
Calculating the correlation function as an $n$-index tensor leads to a generalized Bell inequality.
In this architecture, for an arbitrary $n$-qubit generalized GHZ state, under a set of experimental settings with various angle parameters,
we have obtained  the exact relationship between the amount of violation of the generalized Bell inequality and the variable angle parameters.
By calculating the value of quantum prediction, as a result, we find a set of optimal measurement settings.
Furthermore, as an example, we have shown a simplified description of $n$-qubit system involving two angle parameters.
The main result is that when $l$ is odd there exists a set of optimal measurement settings, $\theta_1=\pi/2$ and $\theta_2=\pi/4$, and otherwise it does not exist.
Finally, we calculate the quantum prediction for the generalized GHZ state with complex coefficients $\alpha$ and $\beta$.  With the modified optimal measurement settings, an important criterion for the violation of the generalized Bell inequality have been demonstrated.
Indeed, it is an interesting and useful fact in experimental tests of multipartite Bell-type inequalities.

\begin{acknowledgements}
This work was supported by the National Natural Science Foundation of China under Grant Nos: 11475054,  11547169,
the Hebei Natural Science Foundation of China under Grant Nos: A2016205145, A2018205125,
the Fundamental Research Funds for the Central Universities of Ministry of Education of China under Grant Nos: 3142017069, 3142015044,
the Foundation for High-Level Talents of Chengde Medical University under Grant No: 201701,
the Research Project of Science and Technology in Higher Education of Hebei Province of China under Grant No: Z2015188.
\end{acknowledgements}

\appendix
\section {}

In order to compute the  quantum prediction
\begin{eqnarray}\label{sum-QC-1}
\sum_{j_1,j_2,\cdots,j_n = 0,1} |q_{j_1,j_2,\cdots,j_n}| =
\frac{|\alpha\beta^{}|}{2^{n-1}} \sum_{j_1,j_2,\cdots,j_n = 0,1} |\sum_{k_1,k_2,\cdots,k_n = 0,1} (-1)^{\textbf{k} \centerdot \textbf{j}}
\text{Re}[\text{e}^{-\text{i}(\phi_1^{k_1}+\phi_2^{k_2}+\cdots+\phi_n^{k_n})}]|,
\end{eqnarray}
we first let
\begin{eqnarray}\label{A-j}
A(\textbf{j}) = \sum_{k_1,k_2,\cdots,k_n = 0,1} (-1)^{\textbf{k}{} \centerdot \textbf{j}} \text{e}^{ -\text{i} (\phi_1^{k_1}+\phi_2^{k_2}+\cdots+\phi_n^{k_n})}.
\end{eqnarray}

A simple calculation shows that
\begin{eqnarray}
A(\textbf{j})
&=& \sum_{k_1,k_2,\cdots,k_n = 0,1} (-1)^{k_1j_1} \text{e}^{ -\text{i} \phi_1^{k_1}}(-1)^{k_2j_2} \text{e}^{ -\text{i} \phi_2^{k_2}}\cdots(-1)^{k_nj_n} \text{e}^{ -\text{i} \phi_n^{k_n}}\nonumber\\
&=& \prod_{l=1,2,\cdots,n}[\text{e}^{-\text {i}\phi_l^0} + (-1)^{j_l}\text{e}^{-\text{i}\phi_l^1}]\nonumber\\
&=& \prod_{l=1,2,\cdots,n}\text{e}^{-\text{i}\frac {(\phi_l^1+\phi_l^0)}{2}}[\text{e}^{\text{i}\frac {(\phi_l^1-\phi_l^0)}{2}} + (-1)^{j_l}\text{e}^{-\text{i}\frac {(\phi_l^1-\phi_l^0)}{2}}]\nonumber\\
&=& \prod_{l=1,2,\cdots,n}\text{e}^{-\text{i}\frac {(\phi_l^1+\phi_l^0)}{2}}2[\text{cos}\frac {(\phi_l^1-\phi_l^0)}{2}]^{1-j_l}[\text{i}\text{sin}\frac {(\phi_l^1-\phi_l^0)}{2}]^{j_l}.
\end{eqnarray}
It follows immediately the definition of $\alpha_l$, $\beta_l$ in Eq.(22) that
\begin{equation}
A(\textbf{j}) = 2^n\text{e}^{\text{i}\sum_{l=1,2,\cdots,n}(-\alpha_l+{\frac{\pi}{2}}j_l)}\prod_{l=1,2,\cdots,n}[\text{cos}\beta_l]^{1-j_l}[\text{sin}\beta_l]^{j_l}.
\end{equation}
Taking the real part of $A(\textbf{j})$, there is
\begin{equation}
\text{Re} [A(\textbf{j})] = 2^n\text{cos}[\sum_{l=1,2,\cdots,n}(-\alpha_l+{\frac{\pi}{2}}j_l)]\prod_{l=1,2,\cdots,n}[\text{cos}\beta_l]^{1-j_l}[\text{sin}\beta_l]^{j_l}.
\end{equation}

By calculation, one derive
\begin{eqnarray}
& & \sum_{\textbf{j}}|\text{Re} [A(\textbf{j})]|\nonumber\\
& =& 2^n\sum_{\textbf{j}}|\text{cos}[\sum_{l=1,2,\cdots,n}(-\alpha_l+{\frac{\pi}{2}}j_l)]|\prod_{l=1,2,\cdots,n}|\text{cos}\beta_l|^{1-j_l}|\text{sin}\beta_l|^{j_l}\nonumber\\
& =& 2^n\sum_{\textbf{j}}|\cos(\sum_{l=1,2,\cdots,n}\alpha_l)\cos(\frac{\pi}{2}\sum_{l=1,2,\cdots,n}j_l)+
\sin(\sum_{l=1,2,\cdots,n}\alpha_l)\sin(\frac{\pi}{2}\sum_{l=1,2,\cdots,n}j_l)|\prod_{l=1,2,\cdots,n}|\text{cos}\beta_l|^{1-j_l}|\text{sin}\beta_l|^{j_l} \nonumber\\
&=&2^n|\sin(\sum_{l=1,2,\cdots,n}\alpha_l)|\sum_{\textbf{j}, \sum_{l=1,2,\cdots,n}j_l=\text{odd}}\prod_{l=1,2,\cdots,n}|\text{cos}\beta_l|^{1-j_l}|\text{sin}\beta_l|^{j_l}
\nonumber\\ & &
+2^n|\cos(\sum_{l=1,2,\cdots,n}\alpha_l)|\sum_{\textbf{j}, \sum_{l=1,2,\cdots,n}j_l=\text{even}}\prod_{l=1,2,\cdots,n}|\text{cos}\beta_l|^{1-j_l}|\text{sin}\beta_l|^{j_l},
\end{eqnarray}
where the following identities
\begin{equation}
 \sum_{l=1,2,\cdots,n}j_l=\text{odd}, ~~~~~ \sin(\frac{\pi}{2}\sum_{l=1,2,\cdots,n}j_l)=\pm 1, ~~~~~~ \cos(\frac{\pi}{2}\sum_{l=1,2,\cdots,n}j_l)=0,\\
\end{equation}
\begin{equation}
 \sum_{l=1,2,\cdots,n}j_l=\text{even}, ~~~~~ \sin(\frac{\pi}{2}\sum_{l=1,2,\cdots,n}j_l)=0, ~~~~~~ \cos(\frac{\pi}{2}\sum_{l=1,2,\cdots,n}j_l)=\pm 1\\
\end{equation}
are used in the last equality.

Then  we obtain
\begin{eqnarray}\label{A-Re-A}
& & \sum_{\textbf{j}}|\text{Re} [A(\textbf{j})]|\nonumber\\
&=&2^{n-1}(|\cos(\sum_{l=1,2,\cdots,n}\alpha_l)|+|\sin(\sum_{l=1,2,\cdots,n}\alpha_l)|)\prod_{l=1,2,\cdots,n}(|\cos\beta_l|+|\sin\beta_l|)\nonumber\\
 & &+2^{n-1}(|\cos(\sum_{l=1,2,\cdots,n}\alpha_l)|-|\sin(\sum_{l=1,2,\cdots,n}\alpha_l)|)\prod_{l=1,2,\cdots,n}(|\cos\beta_l|-|\sin\beta_l|).
\end{eqnarray}
Therefore, the quantum prediction  is given by
 \begin{eqnarray}\nonumber
&&\sum_{j_1,j_2,\cdots,j_n = 0,1} |q_{j_1,j_2,\cdots,j_n}|\nonumber\\
&=&|\alpha\beta|\{[|\cos(\sum_{l=1,2,\cdots,n}\alpha_l)|+|\sin(\sum_{l=1,2,\cdots,n}\alpha_l)|]\prod_{l=1,2,\cdots,n}(|\cos\beta_l|+|\sin\beta_l|)\nonumber\\
 & &+[|\cos(\sum_{l=1,2,\cdots,n}\alpha_l)|-|\sin(\sum_{l=1,2,\cdots,n}\alpha_l)|]\prod_{l=1,2,\cdots,n}(|\cos\beta_l|-|\sin\beta_l|)\},
\end{eqnarray}
as desired.

\section {}

By the definition  $\sum_{l=1,2,\cdots,n}\alpha_l=\beta_{0}$, the quantum prediction
\begin{eqnarray}\nonumber
\sum_{j_1,j_2,\cdots,j_n = 0,1} |q_{j_1,j_2,\cdots,j_n}|=|\alpha\beta|\{\prod_{l=0,1,2,\cdots,n}(|\cos\beta_l|+|\sin\beta_l|)+\prod_{l=0,1,2,\cdots,n}(|\cos\beta_l|-|\sin\beta_l|)\}.
\end{eqnarray}
Obviously, the quantum prediction is the function of $n+1$ independent variables $\beta_{0}, \beta_{1},\cdots, \beta_{n} \in [0,\pi]$ by virtue of the periodic nature of the absolute values of sine and cosine functions.
Furthermore,  one can divide each of  variables $\beta_{0}, \beta_{1},\cdots, \beta_{n} \in [0,\pi]$ into two sections $[0,{\pi}/{2}]$ and $[{\pi}/{2}, \pi]$.

Note that for $\beta_{l} \in [0, {\pi}/{2}]$,
\begin{equation}
|\cos \beta_{l} |+|\sin \beta_{l} |= \sqrt 2 \sin(\beta_{l} +\frac {\pi}{4}),
\end{equation}
\begin{equation}
|\cos \beta_{l} |-|\sin \beta_{l} |= \sqrt 2 \cos(\beta_{l} +\frac {\pi}{4}).
\end{equation}

While for $\beta_{l} \in [{\pi}/{2},\pi]$,
\begin{equation}
|\cos \beta_{l}|+|\sin \beta_{l}|= \sqrt 2 \sin(\beta_{l}-\frac {\pi}{4})= \sqrt 2 \sin[(\beta_{l}-\pi/2)+\frac {\pi}{4}],
\end{equation}
\begin{equation}
|\cos \beta_{l}|-|\sin \beta_{l}|= -\sqrt 2 \cos(\beta_{l}-\frac {\pi}{4})= -\sqrt 2 \cos[(\beta_{l}-\pi/2)+\frac {\pi}{4}].
\end{equation}

 We use $\beta_{l}^0$, $\beta_{l}^1$ to denote $\beta_{l} \in [0, {\pi}/{2}]$,  $\beta_{l} \in [{\pi}/{2},\pi]$, respectively. Thus, the $n+1$ intervals $\beta_l$ in $[0, {\pi}]$,  can be divided into $2^{n+1}$ sections with $\beta_l^0$ in $[0, {\pi}/2]$ and $\beta_l^1$ in $[{\pi}/2, \pi]$.
Therefore,  in the section where the number of  $\beta_l^0$ in $[0, {\pi}/2]$ is  $(n+1-m)$, while the number of  $\beta_l^1$ in $[0, {\pi}/2]$ is $m$, the quantum prediction should be
\begin{eqnarray}\label{aa}
&&\sum_{j_1,j_2,\cdots,j_n = 0,1} |q_{j_1,j_2,\cdots,j_n}|\nonumber\\
&=&|\alpha\beta|
2^{\frac {n+1}{2}}\{\prod_{l=0}^{m-1}\sin[(\beta_{l}^1-\pi/2)+\frac {\pi}{4}]\prod_{l=m}^{n}\sin(\beta_l^0+\frac {\pi}{4})\nonumber\\
&&+(-1)^{m}\prod_{l=0}^{m-1}\cos[(\beta_{l}^1-\pi/2)+\frac {\pi}{4}]\prod_{l=m}^{n}\cos(\beta_l^0+\frac {\pi}{4})\}.
\end{eqnarray}
Obviously, in this section Eq. (\ref{aa}) is equivalent to
\begin{equation}
\sum_{j_1,j_2,\cdots,j_n = 0,1} |q_{j_1,j_2,\cdots,j_n}|=|\alpha\beta|
2^{\frac {n+1}{2}}[
\prod_{l=0}^{n}\sin(\delta_l+\frac {\pi}{4})+(-1)^{m}\prod_{l=0}^{n}\cos(\delta_l+\frac {\pi}{4})], ~~~ \delta_l\in [0, {\pi}/2].
\end{equation}
 Let
 \begin{equation}
 F=|\alpha\beta|2^{\frac {n+1}{2}}[
\prod_{l=0}^{n}\sin(\delta_l+\frac {\pi}{4})+\prod_{l=0}^{n}\cos(\delta_l+\frac {\pi}{4})], ~~~ \delta_l\in [0, {\pi}/2];
\end{equation}
\begin{equation}
 G=|\alpha\beta|2^{\frac {n+1}{2}}[
\prod_{l=0}^{n}\sin(\delta_l+\frac {\pi}{4})-\prod_{l=0}^{n}\cos(\delta_l+\frac {\pi}{4})], ~~~ \delta_l\in [0, {\pi}/2].
\end{equation}

Then,    maximum value of the quantum prediction
\begin{equation}
\textrm{max}\sum_{j_1,j_2,\cdots,j_n = 0,1} |q_{j_1,j_2,\cdots,j_n}|=\textrm{max} \{\textrm{max} F, \textrm{max} G\}.
\end{equation}

Solving sets of differential equations $\partial F/\partial \delta_{l} =0$
produces $\delta_{l}=\pi/4$, $l=0,1,2,\cdots, n$.
With these values it is obvious that the $(n+1)\times (n+1)$ matrix $(\partial^{2} F/{\partial {\delta_{i}}{\partial\delta_{j}}})$ is negative. So we have
$$\textrm{max} F=|\alpha\beta|2^{\frac {n+1}{2}}.$$
Similarly, there is
$$\textrm{max} G=|\alpha\beta|2^{\frac {n+1}{2}},$$
while also occurs at $\delta_{l}=\pi/4$, $l=0,1,2,\cdots, n$, that means $\beta_{l}^0=\pi/4, \beta_{l}^1=\pi/4+\pi/2$.

Together with the periodicity condition, we  arrive at the conclusion that
 the maximum value of the quantum prediction is
\begin{equation}
\textrm{max}\sum_{j_1,j_2,\cdots,j_n = 0,1} |q_{j_1,j_2,\cdots,j_n}|=|\alpha\beta|
2^{\frac {n+1}{2}},
\end{equation}
which  occurs at $\beta_{l}=(2k+1)\pi/4$, $k=0, \pm1,\pm2,\cdots,$ $l=0,1,2,\cdots, n$.

\end{document}